\documentclass[preprint2,epsfig]{aastex}

\newcommand{\lsim}{\stackrel{\scriptstyle <}{\phantom{}_{\sim}}}
\newcommand{\gsim}{\stackrel{\scriptstyle >}{\phantom{}_{\sim}}}
\newcommand{\beq}{\begin{equation}}
\newcommand{\eeq}{\end{equation}}

\slugcomment{MPG-VT-UR 190/99}

\shorttitle{Blaschke, Kl\"ahn and Voskresensky}
\shortauthors{Diquark Condensates and Cooling}

\begin{document}


\title{Diquark Condensates and Compact Star Cooling}


\author{D. Blaschke and  T. Kl{\"a}hn}
\affil{Rostock University, D-18051 Rostock, Germany}
\and
\author{D.~N.~Voskresensky}
\affil{Moscow Institute for Physics and Engineering, 
115409 Moscow, Russia}




\begin{abstract}
The effect of color superconductivity on the cooling of quark stars 
and neutron stars with large quark cores
is investigated.
Various known and new quark-neutrino processes are studied. 
As a result, stars being in the color flavor locked (CFL)
color superconducting phase cool down  
extremely fast. 
Quark stars with no crust 
cool down  too rapidly in disagreement with X-ray data. 
The cooling of stars being in the $N_f =2$
color superconducting (2SC) phase with a crust
is compatible with existing X-ray data.  Also the cooling history
of stars with hypothetic pion condensate nuclei and a crust does not
contradict the data.
\end{abstract}




\section{Introduction}

The interiors of compact stars have been discussed as systems where 
high-density phases of strongly interacting matter do occur in nature, 
see \citep{G96} and \citep{W99}.
The consequences of different phase transition scenarios for the cooling 
behaviour of compact stars have been reviewed recently 
\citep{P92,SWWG96,SHWW97} in comparison with existing X-ray data.

A particular discussion has been devoted to the idea that a strong attraction 
in three flavor $uds$-quark matter may allow for the existence of super-dense 
anomalous nuclei and strange quark stars \citep{B71,W84,RG84}. 
Thereby, in dependence on the value of the bag constant $B$ different possible 
types of stars 
were discussed: ordinary neutron stars (NS) without any quark core, 
neutron stars with quark matter  present only in their deep interiors 
(for somewhat intermediate values of $B$), neutron stars with a large quark 
core (QCNS) and a crust typical for neutron stars, and quark stars (QS) 
with a tiny crust of normal matter and with no crust (both for low $B$ values).
By  QCNS we understand compact stars in which the hadronic shell is rather 
narrow in the sense that it does not essentially affect the cooling which is 
mainly due to the quark core. 
However, this hadronic shell plays the role of an insulating layer between 
quark matter and the normal crust.
By QS we mean compact stars for which the hadronic shell is absent what allows 
only for tiny crusts with maximum densities below the neutron drip density
($\sim 10^{11}$ g/cm$^3$) and masses $M_{\rm cr}\lsim 10^{-5}M_\odot$ 
\citep{AFO86,HBV91}.
Therefore, we suppose that the difference between the models of QS and QCNS 
regarding their cooling behaviour is only in the thickness of the crust. 
This results in completely different relations between
internal ($T$) and surface ($T_{\rm s}$) temperatures for QS and QCNS. 

Recent works 
\citep{ARW98,RSSV98,S98,CD99,RSSV99,BR98,BRS99,PR99,ARW99,SW99a,ABR99,SW99b} 
demonstrate the possibility of 
diquark condensates characterized by large pairing gaps  
($\Delta \sim 100 $ MeV) 
in quark cores of some neutron stars and in  QS and discuss
different possible phases of quark matter. 
Large gaps arise for quark-quark interactions motivated by instantons 
\citep{DFL96,CD99,RSSV99} and by nonperturbative gluon propagators 
\citep{BR98,BRS99,PR99}.  

To be specific in our predictions we will consider models of
the canonical QS and QCNS of 1.4 solar masses ($1.4\, M_\odot$) at a constant 
density.
The constant density profile is actually a very good approximation for QS 
of the mass $M\leq 1.4\, M_\odot$, see \citet{AFO86}.
We consider the model of
QCNS with a crust of the mass $M_{\rm cr}\sim 10^{-1}M_\odot$,
the model of QS with a tiny crust mass 
($M_{\rm cr}\lsim 10^{-5}M_\odot$)
and the model of QS with no crust. 
For QCNS we shall
use the same $T_s/T$ ratio as for ordinary NS, see \citep{T79,M79,ST83}, 
whereas for QS we use somewhat larger values for this ratio, namely 
$T_s =5\cdot 10^{-2}T$ 
for a tiny crust \citep{HBV91} and $T_s =T$ for a
negligible  crust \citep{P91}. 
In the latter case, however, we assume the existence
of black body photon radiation from the surface as for the cases of 
more extended crusts.
We will estimate the cooling of QS and QCNS first in absence of 
color superconductivity and then in presence of  color superconductivity 
for small quark gaps ($\Delta \sim$ 0.1...1 MeV), as suggested 
by \citet{BL84} and  for large gaps 
($\Delta \sim 100$ MeV) as obtained in
Refs. \citep{ARW98,RSSV98,S98,CD99,RSSV99,BR98,PR99,ARW99,SW99a,ABR99,SW99b}.
In the latter case we will consider two phases: 
the color-flavor-locked $uds$-phase \citep{ARW99,SW99a,ABR99,SW99b} and the 
$N_f =2$ color superconducting phase 
\citep{ARW98,RSSV98,S98,BR98,CD99,RSSV99,BRS99,PR99}, in which the $s$-quark 
is absent and the $ud$-diquark condensate selects a direction in color space 
whereby the color charge has to be compensated by the remaining unpaired 
quarks.

Finally, we want to discuss the question whether the hypothesis of a color 
superconducting quark matter phase in compact star interiors is compatible 
with existing X-ray data.  
\section{Normal quark matter}  

A detailed discussion of the neutrino emissivity of quark matter
has first been given by \citet{I82} where the possibility of 
color superconductivity has not been discussed. 
In this work the quark direct Urca
reactions (QDU) $d\rightarrow ue\bar{\nu}$ and $ue\rightarrow d {\nu}$
were suggested as the most efficient processes. 
Their emissivities were estimated as 
\footnote{We note that the numerical factors in the estimates
below are valid only within an order of magnitude.
This is sufficient for the qualitative comparison of cooling scenarios
we present in this work.}
\begin{eqnarray}\label{neutr-DU}
\epsilon^{\rm QDU}_\nu \sim
10^{26}\alpha_s \left(\frac{\rho_b}{\rho_0}\right) Y_e^{1/3}~
T_{9}^6 ~{\rm erg}~{\rm cm}^{-3}~{\rm sec}^{-1},
\end{eqnarray}
where at baryon densities $\rho_b \simeq 2 \rho_0$ the 
strong coupling constant is $\alpha_s \approx 1$ and decreases 
logarithmically at still higher densities \cite[]{KM76}.
The nuclear saturation density is $\rho_0=0.17 {\rm fm}^{-3}$, 
$Y_e =\rho_e /\rho_b$ is the electron fraction, and
$T_9$ is the temperature in units of $10^9$ K. 
The larger the density of the $uds$-system, the smaller is its
electron fraction. For a density $\rho_b \sim 3\rho_0$  one can expect a
rather low electron fraction of strange quark matter $Y_e \sim $10$^{-5}$
\citep{G96} and eq. (\ref{neutr-DU}) yields 
$\epsilon^{\rm QDU} _\nu \sim 10^{25}
~T_{9}^6 ~ {\rm erg} ~{\rm cm}^{-3}~{\rm sec}^{-1}$, see \citep{DSW83,HBV91}.
We did not yet discuss the strange quark contribution given by 
the direct Urca processes $s\rightarrow ue\bar{\nu}$ and 
$ue\rightarrow s {\nu}$. 
Although these processes can occur, their contribution is suppressed compared 
to the corresponding $ud$-reactions \citep{DSW83} by an
extra factor $\mbox{sin}^2 \theta_{\rm C} \sim 10^{-3}$, where 
$\theta_{\rm C}$ is the Cabibbo angle.

If for somewhat larger density the electron fraction was too small 
($Y_e<Y_{ec}\simeq
\sqrt{3}\pi m_e^{3}/(8\alpha_s^{3/2}\rho_b) 
\leq 2\cdot 10^{-8}$, for $\alpha_s \simeq 0.7$ and $\rho_b \simeq 5\rho_0$,
$m_e$ is the electron mass), then all the QDU processes  would be completely
switched off \citep{DSW83} and the neutrino emission would be
governed by two-quark reactions like the quark modified Urca (QMU)
$dq\rightarrow uqe\bar{\nu}$ and the 
quark bremsstrahlung (QB) processes
$q_1 q_2 \rightarrow q_1 q_2 \nu\bar{\nu}$. 
The emissivities of the QMU and QB processes were estimated as \citep{I82}
\begin{eqnarray}\label{neutr-B}
\epsilon^{\rm QMU}_\nu\sim \epsilon^{\rm QB}_\nu ({\rm scr}) \sim 
10^{20}~T_{9}^8 {\rm erg}~{\rm cm}^{-3}~{\rm sec}^{-1}~.
\end{eqnarray} 
The latter estimate of QB emissivity is done in the 
suggestion that the exchanged gluon is screened. If one suggests that
quarks are coupled by transverse non-screened gluons
then one gets \citep{P80}
$\epsilon^{\rm QB}_\nu ({\rm unscr}) \sim 
10^{22}~T_{9}^6 ~{\rm erg}~{\rm cm}^{-3}~{\rm sec}^{-1}$~. 
With a nonperturbative gluon exchange \citep{BR98,BRS99} we expect that the 
estimate 
$\epsilon^{\rm QB}_\nu ({\rm scr})$ is more 
appropriate than the one given by 
$\epsilon^{\rm QB}_\nu ({\rm unscr})  $. 
Therefore, in evaluating the emissivity of 
QB processes we will use (\ref{neutr-B}).
Other neutrino processes (like the plasmon decay 
$\gamma_{\rm pl} \rightarrow ee^{-1}\rightarrow \nu\bar{\nu}$
which goes via coupling to intermediate electron-electron hole 
states, and the corresponding color plasmon decay 
$g_{\rm pl} \rightarrow qq^{-1}\rightarrow \nu\bar{\nu}$ which goes via
coupling of the gluon
to quark-quark hole states \citep{I82}, 
see Fig. 4)
have much smaller emissivities in the normal quark matter phase under 
consideration and can be neglected. 

Among  the processes in the crust,
the electron bremsstrahlung on nuclei gives the largest contribution 
to the emissivity as estimated in \citep{I82}
\begin{eqnarray}\label{neutr-cr}
\epsilon^{\rm cr}_\nu \sim 10^{21} \left(\frac{M_{cr}}{M_\odot}\right)~
T_{9}^6~{\rm erg}~{\rm cm}^{-3}~{\rm sec}^{-1}~.
\end{eqnarray} 
This contribution can be neglected for QS  
due to a tiny mass of the QS crust $M_{cr}\lsim 10^{-5}M_\odot$.

Besides, one should add the photon contribution
\begin{eqnarray}\label{photon}
\epsilon_{\gamma} \simeq 2\cdot 10^{18}\left(\frac{R}{10~{\rm km}}\right)^2~
T_{s7}^4~{\rm erg}~{\rm cm}^{-3}~{\rm sec}^{-1}, 
\end{eqnarray} 
where $T_{s7}$ is the surface temperature in units of $10^7$ K, 
see \citep{ST83}. 
This process becomes the dominant one for QS at essentially shorter times 
than for the QCNS due to the higher $T_s /T$ ratios for the former.  

Internal and surface temperatures are related by a coefficient determined by 
the scattering processes occurring in the outer region, where the electrons 
become non-degenerate. 
For NS with rather thick crust, an appropriate fit to numerical calculations 
\citep{T79} is given by a simple formula \citep{ST83}
\begin{eqnarray}\label{temp-rel}
T_s =(10~T)^{2/3}~,
\end{eqnarray}
where $T_s$ and $T$ both are measured in units of K.
We shall use this expression when dealing with QCNS.  
A rough estimate of (\ref{temp-rel}) yields
$T_s =a\times 10^{-2} T$ with $a\simeq 0.2 - 2$ in dependence on the value 
of the internal temperature varying in the interval $10^{10} \dots 10^{7}$ K
of our interest.
In the QS the crust is much more thin than in the NS 
and the ratio $T_s / T$ should be
significantly larger. 
Therefore, depending on the thickness of the crust we shall use two estimates
for QS scenarios:  $T_s =5\times 10^{-2} T$
for a tiny crust \citep{HBV91}, 
and  $T_s \simeq T$ for negligible  crust \citep{P91}. 

In order to compute the cooling history of the star 
we still need the specific heat of the 
electron, photon, gluon and quark sub-systems. 
In accordance with the estimates of \citet{I82}, \citet{HBV91} we have 
\begin{equation}\label{el}
c_{e}\simeq 0.6 \cdot  10^{20}
\left(\frac{Y_e\rho_b}{\rho_0}\right)^{2/3} T_9 
~{\rm erg}~{\rm cm}^{-3}~{\rm K}^{-1}, \\
\end{equation}
\begin{equation}\label{gam-g}
c_{\gamma}\simeq 0.3\cdot 10^{14}~T_{9}^{3}
~{\rm erg}~{\rm cm}^{-3}~{\rm K}^{-1}~, \\
\end{equation}
\begin{equation}\label{gluon}
c_{g}\simeq 0.3 \cdot 10^{14}~ N_g T_{9}^{3}
~{\rm erg}~{\rm cm}^{-3}~{\rm K}^{-1}~, \\
\end{equation}
\begin{equation}\label{quark}
c_{q}\simeq 10^{21} \left(\frac{\rho_b}{\rho_0}\right)^{2/3} T_9
~{\rm erg}~{\rm cm}^{-3}~{\rm K}^{-1}~, 
\end{equation}
where $N_g$ is the number of different color states of
massless gluons.
The very small contribution to the specific heat
of the crust can be neglected \citep{LRPP94}.
The cooling equation reads   
\begin{eqnarray}\label{sum}
\sum_{i=q, e,\gamma , g} c_i \cdot\frac{d T}{dt}= 
- \epsilon_\gamma~~ - \sum_{j={\rm QDU, QMU, QB, cr}} \epsilon_\nu^j  ,
\end{eqnarray}
where the summation is over all contributions to the specific heats 
and emissivities as discussed above. 
The evolution at large times on which we are focusing our interest here
is insensitive to the assumed value of the initial temperature $T_0$. 
We checked it using different values of initial temperature.
To be specific we choose the value $T_{0,9} =10$ as a typical initial 
temperature for proto-neutron stars.

In Figs. \ref{fig1}-\ref{fig3} we show the cooling history of QCNS with 
standard thickness of the NS crust (when internal and surface temperatures 
are related by Eq. (\ref{temp-rel})), QS with a tiny crust 
($T_s =5\cdot 10^{-2}T$) and QS with negligible crust ($T_s =T$), respectively.
Solid curves are for the matter suggested to be in the normal state, i. e. 
in the absence of color superconductivity. 
Different groups of data points are taken from Table 3 of Ref. \citep{SSWW99}
where the notations are explained. 
In the lower panels of Figs. \ref{fig1}-\ref{fig3}, we show the results for 
$Y_e >Y_{ec}$ taking $Y_e =10^{-5}$, $\alpha_s =1$, $\rho = 3\rho_0$. 
This is a representative set of parameters for
which the QDU processes contribute to the cooling.
We see that two low-temperature pulsars can be explained as QCNS being in 
normal state (Fig. 1, thick solid curve, lower panel)
and many observations can be interpreted as QS  in normal state
with a tiny crust (Fig. 2, lower panel). 
 
The upper panels of Figs. 1-3 demonstrate the cooling history of QCNS and QS
for $Y_e <Y_{ec} $, namely for $Y_e =0$, $\alpha_s =0.7$, and $\rho = 
5\rho_0$. QCNS  being in normal state (Fig. 1, thick solid curve, 
upper panel) cool down rather slowly but are still in agreement with the 
data for a few pulsars.
For QS with a tiny crust (Fig. 2, $T_e =5\times 10^{-2} T$) we get a nice 
fit of many data points. 

In both cases $Y_e <Y_{ec} $ and $Y_e >Y_{ec}$, see Fig. 3, QS 
with negligible crust cool down too fast in disagreement with the X-ray data.

\section{ Color superconductivity} 
In the standard scenario of NS cooling the inclusion of
nucleon pairing suppresses the emissivity resulting in a more 
moderate cooling. 
Now, considering QS and QCNS we will show that we deal with the opposite case.

Due to the pairing, the emissivity of QDU processes
is suppressed by a factor $\mbox{exp}(-\Delta /T)$
and the emissivities of QMU and QB processes are suppressed by a 
factor $\mbox{exp}(-2\Delta /T)$ for $T<T_c$. Thereby in our
calculations we will use expression (\ref{neutr-DU}) for QDU suppressing
the rate by 
$\mbox{exp}(-\Delta /T)$ and expressions (\ref{neutr-B}) for QMU and QB 
suppressing
the rates
by $\mbox{exp}(-2\Delta /T)$ for $T<T_c$.
We also observe that plasmon and color plasmon decay processes are switched off
in the superconducting phase when the photons and the gluons acquire masses  
due to the Higgs effect, as it happens for photons in usual superconductors. 
\citet{VKK98} however demonstrated that in superconducting 
matter there appears a new neutrino neutral current process analogous to the 
plasmon decay but now due to a massive photon decay.  
Its emissivity is suppressed by the factor
$\mbox{exp}(-m_{\gamma}/T)$ rather than by $\mbox{exp}(-\Delta/T)$,
as for direct Urca processes, or by $\mbox{exp}(-2\Delta/T)$, 
as for modified Urca and corresponding bremsstrahlung processes.
This results in a large contribution for small but finite values of 
$m_{\gamma}$.
Naively, one could expect that in a color superconductor
the squared photon mass is  proportional to the 
fine structure constant $\alpha =1/137$ as it is in ordinary
superconductors.  
Then it would be much smaller than the squared 
gluon mass since the latter quantity in the color superconducting phase
has to be proportional to the corresponding strong coupling constant 
$\alpha_s$. 
In reality, due to the common gauge transformation for 
electromagnetic and color fields one deals with mixed 
photon-gluon excitations. We demonstrate this 
using the expression for the free energy density of the
color superconducting phase \citep{BL84}, 
\begin{eqnarray}
\label{gauge}
f&=&f_n + a d^{\ast}d +\frac{1}{2}b (d^{\ast}d)^2 \nonumber \\
&+&c (\nabla +iq\vec{A}+\frac{ig\vec{B}}{\sqrt{3}})d^{\ast}
(\nabla -iq\vec{A}-\frac{ig\vec{B}}{\sqrt{3}})d \nonumber \\
&+&\frac{(\mbox{rot}\vec{A})^2}{8\pi}+ \frac{(\mbox{rot}\vec{B})^2}{8\pi},
\end{eqnarray}
where $f_n$ is the normal part of the free energy density, $a =\mu p_{Fq}
t/\pi^2$, $t=(T-T_c)/T_c <0$, $b =7\zeta (3) \mu p_{Fq}/(8\pi^4 T_c^{2})$,
$c =p_{Fq}^2 b /(6\mu^2)$, $\mu\simeq p_{Fq}$ is chemical potential
of ultra-relativistic quark, $q =\sqrt{\alpha}/3$ is the 
electric charge of a $ud$-pair, $\alpha =1/137$.
We have introduced an interaction with two gauge fields $A_\mu$
and $B_\mu$. $A_\mu$ is the electromagnetic field and $B_\mu$
relates to the gluons.
For simplicity we consider only fluctuations of space-like components of the 
fields and assume the $B_\mu$ field to be an Abelian field. 
Variation of (\ref{gauge}) with respect to the fields 
gives the corresponding equations of motion. 
Taking $d=d_0 +d^{\,\prime}$, where 
$d_0 =\sqrt{-a/b}$ is the order parameter and $d^{\, \prime}$ 
is the fluctuating field,
we linearize the equations of motion for the fluctuating fields 
$d^{\, \prime},\vec{A}, \vec{B}$. 
Solving these equations in the Fourier representation, we get three
branches of the spectrum. The branch 
$\omega^2 =\vec{q}^2 +2|\alpha |$ corresponds to 
fluctuations of the order parameter characterized by a large mass 
$m_{d}=\sqrt{-2t\mu p_{Fq}} /\pi \sim m_{\pi}=140$ MeV.  
The branch 
\begin{eqnarray}\label{ph-g}
\omega^2 =\vec{ q}^{\,\, 2} +m_{\gamma ,g}^2
\end{eqnarray}
describes a massive photon-gluon excitation with a mass $m_{\gamma ,g}^2 =
8\pi c (\alpha +
3\alpha_s )d_0^{2}/9 $.
The extra branch  $\omega^2 =\vec{ q}^{\,\, 2}$ describes a massless mixed 
photon-gluon Goldstone excitation. 

Thus  in difference with a usual proton superconducting phase of NS
where the photon has rather low mass 
$m_{\gamma}= d_0 \sqrt{8\pi c \alpha }\simeq 4$ MeV for $\mu \simeq 400$ MeV,
in the color superconductor we, probably, deal with a much more massive
mixed photon-gluon excitation
(with $\alpha$ being replaced by  $\alpha + 3\alpha_s$) 
and with the corresponding Goldstone boson\footnote{However, the penetration 
depth of the external magnetic field is 
associated with the above mentioned small photon mass rather 
than with the massive photon-gluon excitation
or with the massless Goldstone excitation, see \citep{BSS99}.}. 

The Goldstone boson does not contribute to
the mentioned photon-gluon decay process, whereas the massive 
excitation does. Now, armed with an expression for the photon-gluon mass we
may estimate the emissivity of the corresponding processes
$(\gamma -g) \rightarrow ee^{-1}+qq^{-1}\rightarrow \nu\bar{\nu}$, where
$e^{-1}$ and $q^{-1}$ are the 
electron hole and the quark hole, respectively, see Fig. 4. Using
the result for $\gamma \rightarrow ee^{-1}+pp^{-1}
\rightarrow \nu\bar{\nu}$ \citep{VKK98} we easily get
\begin{eqnarray}\label{em-ph}
\epsilon_{(\gamma -g)}&\sim& 10^{29} \left( \frac{m_{\gamma ,g}}{\mbox{MeV}}
\right)^{7/2}T_9^{3/2}\left( 1+\frac{3T}{2m_{\gamma ,g}}\right)
\nonumber \\
&\times&\mbox{exp}(-m_{\gamma,g}/T)~~\mbox{erg}~\mbox{cm}^{-3}~\mbox{sec}^{-1},
 \end{eqnarray}
for $T<m_{\gamma ,g}$, and using the condition $\Delta \ll \mu$.
As we see, the emissivity of this process  is strongly 
suppressed for the values 
$m_{\gamma ,g}\simeq 70$ MeV following from eq. (\ref{ph-g}). 
Also the specific heat of this mixed photon-gluon excitation is suppressed by 
the same exponential factor $\mbox{exp}(-m_{\gamma ,g}/T)$.
For the Goldstone excitation the contribution to the specific heat is given by 
Eq. (\ref{gam-g}). 
 
For the quark specific heat at $T<T_c$ we use an expression similar to the one 
which applies for the case of nucleon pairing \citep{M59,M79,HBV91}, i.e.
\begin{eqnarray}\label{heat}
c_{sq}&=&3.2~c_{q}~(T_c /T)~\mbox{exp}(-\Delta/T)
\nonumber \\&\times& 
\left[ 2.5-1.7~T/T_c +3.6~(T/T_c)^2 \right],
\end{eqnarray}
where $T_c$ is related to $\Delta$ as $\Delta =1.76~ T_c$ for the case
of small gaps. 
For CFL and 2SC phases we will use $T_c \simeq 0.4 ~\Delta $ .
Actually, in the latter case the relation between  $T_c$ and $\Delta$ is 
unsettled.
However, one believes that the coefficient in standard BCS formula
$T_c \simeq 0.57~\Delta $ is appreciably reduced as impact of 
instanton-anti-instanton molecules \citep{RSSV99}. 
The mentioned uncertainty in the value of $T_c$ for CFL and 2SC phases
does not significantly affect the cooling curves since the dominant 
effect comes from the exponential factor where $\Delta$ enters rather than 
$T_c$.

Now, armed with all necessary expressions we may estimate the cooling of QS and
QCNS being in $uds$- or 2SC phases (except for the 
crust). 

\subsection{Cooling of different $uds$-phases: $uds$-small gaps, CFL~($Y_e 
>Y_{ec}$), and
CFL~($Y_e <Y_{ec}$)} 
We select the following values of the pairing gaps: small gaps 
$\Delta = 0.1 \dots 1$ MeV  as suggested to occur in the color superconducting 
region by \citet{BL84} 
and a large gap $\Delta \sim 50$ MeV, 
as suggested for CFL phase in recent works \citep{ARW98,RSSV99}.
\citet{HBV91} have discussed the cooling of superconducting QS and QCNS for 
very small gaps with critical temperatures $T_{c9} = 0.1, 0.5, 1$. 
However, for QDU processes the suppression factor $\mbox{exp}(-2\Delta /T)$  
has been used rather than $\mbox{exp}(-\Delta /T)$. 
The value of the critical temperature $T_{c9} = 0.1$ seems to be quite small, 
therefore we use $\Delta =0.1 \dots 1$ MeV for the case of small gaps. 

We calculate the cooling history of QS and QCNS using eq. (\ref{sum}),
where now summation over $j$ implies summation of emissivities of QDU,
eq. (\ref{neutr-DU}), suppressed by $\mbox{exp}(-\Delta /T)$, QMU and QB, 
eq.(\ref{neutr-B}), suppressed by $\mbox{exp}(-2\Delta /T)$, emissivity
of the crust,
eq. (\ref{neutr-cr}), and emissivity of photon-gluon decay, eq. (\ref{em-ph}).
Summation over $i$ implies summation of the quark contribution evaluated 
according to eq. (\ref{heat}), the electron contribution, eq. (\ref{el}), the 
massless photon-gluon Goldstone contribution which coincides with that given 
by eq. (\ref{gam-g}) and the contributions of massive gluons given by eq. 
(\ref{gluon}) suppressed by $\mbox{exp}(-m_{\gamma-g} /T)$ and thus being very
tiny. 
The contribution of the crust to the specific heat is negligible. 
Also in the CFL phase exist 9 hadronic quasi-Goldstone modes. 
Although their masses are not known we may roughly estimate them as
$m_{h}>m_q$, where $m_q$ is the bare quark mass with a minimum value of $\sim
5$~MeV. With these masses the contribution of hadronic quasi-Goldstone modes
is also very small at temperatures of our interest and can be neglected.

The dotted curves in Figs. 1 - 3  demonstrate the cooling history of QCNS 
and QS for the gap $\Delta =0.1$ MeV, whereas the dashed lines correspond to 
the cooling of the CFL phase for $\Delta =50$ MeV. 
All thin dotted and dashed lines correspond to the case when the process of
massive mixed photon-gluon decay is artificially excluded whereas the
corresponding
thick lines 
represent the cooling history when this process is taken into account 
according to Eq. (\ref{em-ph}). 
This new process essentially influences on 
the early stage of the cooling although the mass of the mixed photon-gluon 
excitation was supposed to be very high 
($m_{\gamma, g}=70$ MeV for $Y_e =10^{-5}> Y_{ec}$, lower panel, 
and $m_{\gamma, g}=60$ MeV for $Y_e =0$, upper panel). This is due to to
a big numerical factor in eq. (\ref{em-ph}). 

In all the cases we obtain very rapid cooling in 
disagreement with the data. 
Particularly rapid cooling occurs for the CFL phase.
In the latter case contributions of the QDU, QMU and QB processes to the
emissivity are 
suppressed as well as the quark contribution to the specific heat.
The 
rate is governed by the photon emissivity from the surface. 
For  
$Y_e > Y_{ec}$, the specific heat is determined by the electrons.
As the consequence of this reduction of the specific heat we get a very rapid 
cooling of the CFL ($Y_e > Y_{ec}$) phase, see lower panel of Figs. 1 - 3.
For $Y_e =0$ (upper panel of Figs. 1 - 3) there are no electrons and
the specific heat is determined by a very small 
contribution of the Goldstone mode given by eq. (\ref{gam-g}), 
so that both 
QCNS and QS cool down even faster than for the case 
$Y_e =10^{-5}> Y_{ec}$. 
In both $Y_e > Y_{ec}$ and $Y_e < Y_{ec}$ cases
the cooling time of the CFL phase is extremely small.
This means that in reality the cooling is governed by the heat transport in 
the thin crust \citep{P91}, which we did not take into account.

Thus we see that QCNS and QS, if present among the objects measured in X-rays, 
can't be in the CFL phase.
The cooling of this phase is so rapid that one might expect problems 
not only for the models of QS and QCNS but also for the models of
NS with quark cores consisting of the CFL phase only in deep interiors. 
          
\subsection{Cooling of the 2SC phase} 

This phase is probably more reliable for QCNS rather than for QS since the CFL 
phase is energetically favorable in the latter case. 
The 2SC phase is characterized by large gaps, $\Delta \sim 100$ MeV  
\citep{ARW98,RSSV98,RSSV99}. 
To be specific we suppose that blue-green  and green-blue
$ud$-quarks are paired, whereas red $u$- and $d$-quarks ($u_r, d_r$) remain 
unpaired.
This has the consequence that the QDU processes  
on the red (unpaired) quarks, as $d_r\rightarrow u_r e\bar{\nu}$, as well as 
QMU, 
$d_r q_r\rightarrow u_r q_r e\bar{\nu}$, and QB, $q_{1r} q_{2r}\rightarrow
q_{1r} q_{2r}\bar{\nu}\nu$,
are not blocked by the gaps whereas other processes involving paired
quarks
are blocked out by large diquark gaps. 
The QDU  process on red quarks occurs in the $Y_e >Y_{ec}$ case only. 
Its emissivity is given by  
\begin{eqnarray}\label{DU-super}
\epsilon^{{\rm QDU}}_\nu (d_r )&\sim& 10^{25}\alpha_s~({\rho_b}/{\rho_0})~ 
Y_e^{1/3}~T_{9}^6 \nonumber\\
&& {\rm erg}~{\rm cm}^{-3}~{\rm sec}^{-1}.
\end{eqnarray}
The extra suppression factor of the rate (\ref{neutr-DU}) comes from the fact 
that the number of available unpaired color states is reduced.

QMU and QB processes on red quarks are also rather efficient.
Although there is no one-gluon exchange between $d_r - d_r$, the QMU and
QB processes 
may go via a residual quark-quark
interaction, e.g, via two-gluon exchange. 
We roughly estimate the corresponding 
emissivities as
\begin{eqnarray}\label{neutr-r}
\epsilon^{\rm QMU}_\nu (d_r q_r) &\sim&\epsilon^{\rm QB}_\nu (q_{1r} q_{2r}) 
\nonumber\\
&\sim& 10^{19} T_{9}^8 \,\, \mbox{erg} \, \mbox{cm}^{-3} \, \mbox{sec}^{-1}.
\end{eqnarray}  
In the 2SC ($Y_e >Y_{ec}$) phase the QDU process on red quarks is the
dominant process and QMU and QB processes on red quarks are
subdominant processes whereas in the 2SC ($Y_e <Y_{ec}$) phase
QDU processes do not occur and QMU and QB processes on red quarks become
the dominant processes.
Other processes like QDU, QMU and QB 
with participation of other color and flavor quarks
are continued to be appreciably suppressed by large gaps. 

The specific heat is also changed in the 2SC phase since the
$d_r$ and $u_r$  contributions are not
suppressed by a factor $\mbox{exp}(-\Delta /T)$ whereas color-paired
$ud$-contributions remain to be suppressed. 
With these findings we calculate the cooling history of QCNS and QS. 
The results are presented in Fig. 5
for $Y_e =10^{-5}$, $\rho =3\rho_0$ (thick lines), and  
$Y_e =0$, $\rho =5\rho_0$ (thin lines). 
We see that in both cases the cooling history of 
QCNS and also of QS with a tiny crust ($T_s =5\cdot 10^{-2}T$) 
nicely agrees with the X-ray data. 
The cooling of QS with negligible crust does not agree with the data. 
\section{Conclusions}
We have estimated the contributions of various quark processes to the 
emissivity. 
Among them, the new decay process of the massive mixed photon-gluon excitation 
is operating at the early stage of the cooling and QDU, QMU and QB
processes on red quarks determine the cooling of the 2SC phase.
We discussed the cooling history of QS and QCNS taking into account 
different possibilities: $Y_e >Y_{ec}$ and $Y_e <Y_{ec}$,
the normal quark phase, and various
color superconducting phases as the ``$uds$-phase - small gaps'' suggested 
by \citet{BL84}, the CFL phase, and the 2SC phase, as suggested in recent 
works  \citep{ARW98,RSSV98,S98,ARW99,RSSV99}. 
In all the cases we see that 
{\em QS and QCNS being in the CFL phase cool down  extremely
fast  in disagreement with known X-ray data.}
Also the cooling curves for the case of small gaps ($\Delta = 0.1 \dots 1$ MeV)
disagree with the data. 

Even if the CFL phase would be realised only in the deep interior region of a 
NS it would be problematic to satisfy X-ray data.
In this case the star would radiate mostly not from the surface but from the
CFL region due to its extremely small specific heat related to the Goldstone
excitation. 
Thus the cooling time would be determined by the heat transport from exterior 
regions to the center rather than by the cooling of the hadronic shell.

{\em The cooling history of the QS and QCNS with a crust being in the normal 
state agrees with the data.} 

In this respect the following remark is in order. 
It is now believed that quark matter below $T_c \sim 50$ MeV is in the color 
superconducting state characterized by a diquark condensate with large energy 
gaps ($\Delta \sim 100$ MeV) rather than being in the normal state or the 
superfluid state characterized by small gaps ($\Delta \lsim 1$ MeV). 
If so, one could think that our above discussion of normal quark matter and 
of the case of a small gap has just pedagogic reasoning. 
However, this is not really so. 
Indeed, besides the idea of abnormal strange nuclei and strange stars
\citep{B71,W84,RG84} there is the very similar idea of abnormal
pion condensate nuclei and stars with pion condensate nuclei, 
see \citep{M71,VSC77} and the review \citep{MSTV90}, chapters 15, 16.
The same relates to the kaon condensate objects. 
Pion  condensate systems cool down at about the same rate as given by QDU 
processes (for $Y_e \sim 10^{-5}$) 
\footnote{One should bear in mind a strong suppression
of emissivities of pion condensate 
processes due to nucleon-nucleon correlation effects \citep{BRSSV95} and 
an enhancement of the specific heat \citep{VS84,VS86,MSTV90} 
which are often ignored in the cooling simulations.}.
Besides, they can be in the normal state or in the 
superfluid state characterized by very small gaps $\Delta \lsim 0.1$ MeV. 
The cooling history of systems being in normal state is described by thick 
solid curves on the lower panels of Figs. 1 - 3.
Thus we may also conclude that 
{\em the hypothesis of pion condensate nuclei-stars} 
(being in normal $\Delta =0$ state with a crust) 
{\em does not contradict to the X-ray observations.}    
Stars with pion and kaon condensate nuclei being in the superfluid state
with gaps $\Delta \gsim 0.1$ MeV are ruled out as objects being observed
in X-rays.

{\em The cooling history of the 2SC phase of QCNS and QS with a tiny crust 
($T_s =5\cdot 10^{-2}T$) agrees with the X-ray data.}
{\em The cooling history of QS with no crust disagrees with X-ray data.}

Three final remarks are in order:

(i) It is conceivable that there are more complex
collective effects which essentially affect the specific heat and 
the luminosity. 
E.g., we calculated the mixed photon-gluon spectrum in a 
simplified model of two Abelian gauge fields and concluded that 
the mass of the excitation is large, whereas one can't exclude that in the 
realistic non-Abelian case there exists a photon-gluon excitation of a 
small mass that could lead to very efficient cooling via the mixed photon-gluon
decay process given by eq. (\ref{em-ph}). 
The masses of hadronic quasi-Goldstone modes in CFL phase should be
carefully studied. 

(ii) As we mentioned above in the discussion of the DU process, we have 
neglected the contribution of strange quarks (QDU-s) relative to that of the
light quarks.
The former contribution is $\sim 10^{-3}$ times smaller than the latter in the
case of normal matter and in the case when all diquark gaps are identical, the 
discussion given for the CFL phase applies.
However, if the pairing gaps for strange diquarks are smaller than those for 
nonstrange diquarks the QDU-s contribution to the emissivity can be essentially
enhanced. 
Up to now there exist only rough estimates of the values of the gaps and this 
gives rise to large uncertainties in final estimates of the emissivity.
Above, in order to be specific, we made calculations considering QDU on u- and 
d- quarks only. 
The inclusion of QDU-s in the case when the strange diquark 
gaps can be smaller than those for non-strange diquarks is simulated by varying
the values of the gaps in a wide interval.
This does not change the qualitative picture of the QS and QCNS cooling we 
have discussed in the present work.

(iii) We also would like to point out that, if the compact object
formed in the explosion of SN 1987A was a QS or a QCNS being in the CFL phase 
($Y_e <Y_{ec}$), it is now so cold that it is already impossible to observe it
in soft X-rays.
This becomes particularly interesting if continued observation of SN 1987A
would find a pulsar and would not observe it in X-rays.



\acknowledgments

We acknowledge the important remarks and fruitful discussions by M. Alford and 
K. Rajagopal after reading the draft of the paper. 
We also thank B. Friman, J. E. Horvath, E. Kolomeitsev, R. Pisarski, 
D. Rischke, A. Sedrakian, and F. Weber for their discussions. 
One of us (DNV) is grateful for the hospitality extended to him during the 
visit at Rostock University and acknowledges financial support from the 
{\sc Max-Planck-Gesellschaft}.

\clearpage



\figcaption[fig1.eps]{Cooling history of QCNS  $M=1.4\,M_\odot$. 
The relation between $T_s$ and $T$ is given by eq. 
(\ref{temp-rel}). Lower panel: $Y_e =10^{-5}$, $\rho =3\rho_0$, and
$\alpha_s =1$; upper panel: $Y_e =0$, $\rho =5\rho_0$, and
$\alpha_s =0.7$. 
For the thin dotted and dashed lines the process of 
massive mixed photon-gluon decay is artifically excluded; the 
corresponding thick lines take this process into account. 
Line styles are related to different values of the gap $\Delta$,
given in the legend. Data points are taken from analysis \protect\cite{SSWW99}.
\label{fig1}}

\figcaption[fig2.eps]{Same as in Fig. \ref{fig1} but for QS with 
$T_s =5\cdot 10^{-2}T$.  \label{fig2}}

\figcaption[fig3.eps]{Same as in Fig. \ref{fig1} but for QS with 
$T_s =T$. \label{fig3}}

\figcaption[fig4.eps]{Photon- gluon decay via intermediate quark- quark 
hole and electron- electron hole states. \label{fig4}}

\figcaption[fig5.eps]{Cooling history of QCNS and QS
being in 2SC phase, $M=1.4\,M_\odot$, $\Delta =100$ MeV. 
Thick lines: $Y_e =10^{-5}$, $\rho =3\rho_0$, and
$\alpha_s =1$; thin lines: $Y_e =0$, $\rho =5\rho_0$, and
$\alpha_s =0.7$. Solid curves correspond to QCNS, 
the relation between $T_s$ and $T$ given by eq. 
(\ref{temp-rel}), dashed lines correspond to QS with a tiny crust,
$T_s =5\cdot 10^{-2}T$, and dotted ones to QS without crust, $T_s =T$. 
\label{fig5}}





\end{document}